\documentclass{article}
\usepackage{PRIMEarxiv}

\usepackage[utf8]{inputenc} 
\usepackage[T1]{fontenc}    
\usepackage{hyperref}       
\usepackage{cite}  
\usepackage{booktabs}     
\usepackage{amsfonts}       
\usepackage{nicefrac}       %
\usepackage{microtype}      

\usepackage{fancyhdr}       
\usepackage{graphicx}       
\usepackage{amsmath} 

\usepackage{url}            
\usepackage{booktabs}       
\usepackage{amsfonts}       
\usepackage{nicefrac}       
\usepackage{microtype}      
\usepackage{lipsum}
\usepackage{graphicx}
\usepackage{bm}
\usepackage{multirow}

\usepackage[ruled,linesnumbered]{algorithm2e}

\graphicspath{{media/}}     

\usepackage{xcolor}

\newcommand{\ie}[1]{\textit{i.e.} #1}
\newcommand{\eg}[1]{\textit{e.g.} #1}

\usepackage[mathlines]{lineno}

\title{
Multi-objective optimization via evolutionary algorithm (MOVEA) for high-definition transcranial electrical stimulation of the human brain
\thanks{\textit{\underline{Citation}}: 
\textbf{Wang et al. Title. Pages... DOI:000000/11111.}} 
}

\author{
  Mo Wang$^{1}$, Kexin Lou$^{1,2}$, Zeming Liu$^{1}$, Pengfei Wei$^{3}$, Quanying Liu$^{1}$ \\ \\
  $^{1}$ Department of Biomedical Engineering, 
  Southern University of Science and Technology \\
  $^{2}$ School of Information Technology and Electrical Engineering, University of Queensland \\
  $^{3}$ Shenzhen Institute of Advanced Technology, Chinese Academy of Sciences\\ \\
  Corresponding to Q.L. with \texttt{liuqy@sustech.edu.cn} \\
}

\begin{document}
\maketitle

\begin{abstract}
Designing a transcranial electrical stimulation (TES) strategy requires considering multiple objectives, such as intensity in the target area, focality, stimulation depth, and avoidance zone, which are often mutually exclusive. A computational framework for optimizing different strategies and comparing trade-offs between these objectives is currently lacking. In this paper, we propose a general framework called multi-objective optimization via evolutionary algorithms (MOVEA) to address the non-convex optimization problem in designing TES strategies without predefined direction. MOVEA enables simultaneous optimization of multiple targets through Pareto optimization, generating a Pareto front after a single run without manual weight adjustment and allowing easy expansion to more targets. This Pareto front consists of optimal solutions that meet various requirements while respecting trade-off relationships between conflicting objectives such as intensity and focality. MOVEA is versatile and suitable for both transcranial alternating current stimulation (tACS) and transcranial temporal interference stimulation (tTIS) based on high definition (HD) and two-pair systems. We performed a comprehensive comparison between tACS and tTIS in terms of intensity, focality, and steerability for targets at different depths. Our findings reveal that tTIS enhances focality by reducing activated volume outside the target by 60\%. HD-tTIS and HD-tDCS can achieve equivalent maximum intensities, surpassing those of two-pair tTIS, such as $0.51V/m$ under HD-tACS/HD-tTIS and $0.42V/m$ under two-pair tTIS for the motor area as a target. Analysis of variance in eight subjects highlights individual differences in both optimal stimulation policies and outcomes for tACS and tTIS, emphasizing the need for personalized stimulation protocols. These findings provide guidance for designing appropriate stimulation strategies for tACS and tTIS. MOVEA facilitates the optimization of TES based on specific objectives and constraints, advancing tTIS and tACS-based neuromodulation in understanding the causal relationship between brain regions and cognitive functions and in treating diseases. The code for MOVEA is available at \url{https://github.com/ncclabsustech/MOVEA}.
\end{abstract}


\newpage
\section{Introduction}

Transcranial electrical stimulation (tES) is a non-invasive neuromodulation technique with substantial potential for clinical applications, such as stroke treatment \cite{lefebvre2015neural} and motor function improvement \cite{feurra2011frequency}. By injecting current through electrode pairs on the scalp surface, tES can modulate neural activities and avoid complications from intracranial stimulations. 
tES has many specific implementations, among which transcranial direct current stimulation (tDCS) and transcranial alternating current stimulation (tACS) are the most commonly used in neuromodulation.
tDCS delivers direct current, while tACS delivers low-frequency alternating current, both in the range of 1--2 mA to the brain.
Some studies have reported that tES can improve behavioral performance in cognitive tasks. For example, $10Hz$ tACS over the prefrontal cortex can enhance phonological word decision making~\cite{moliadze2019after}, and $18Hz$ tACS over the right parietal lobe can significantly improve the visual processing~\cite{battaglini2020parietal}.
However, due to the current diffusion effect, the current injected from the scalp by these traditional tES methods may not effectively penetrate the skull and generate focal stimulation in the deep brain region~\cite{datta2009gyri}.

To address these challenges, efforts have been directed toward developing novel electrode configurations and electrical stimulation technologies. For the former, Bortoletto et al. have proposed high-definition transcranial electrical stimulation (HD-tES)~\cite{Bortoletto2016ReducedCS}. By replacing large sponge electrodes with a set of small saline electrodes, the high-definition electrode configuration in HD-tES helps generate a more convergent electrical potential field. For the latter, Grossman et al. have proposed a new stimulation method, namely transcranial temporal interfering stimulation (tTIS)~\cite{grossman2017noninvasive}. tTIS applies two electrode pairs to introduce two high-frequency carriers with a small frequency difference, resulting in a low-frequency envelope at the target by coupling the carriers.
For example, two pairs of electrodes with $2000 Hz$ and $2010 Hz$ transcranial electrical stimulations can create a $10 Hz$ envelope at the specific target region. Due to nerves being less sensitive to high-frequency stimulation (\eg 1 $KHz$)~\cite{hutcheon2000resonance}, the high-frequency carriers do no activate the neuronal tissues outside the target region. This effect helps limit effective stimulation to the low-frequency envelope zone, creating a more focal stimulation in tTIS. 
To improve the stimulation effects, Huang et al. and Cao et al. have relieved the limitation of two-pair  electrodes to multi-pair electrode-based tTIS~\cite{cao2019stimulus} and high-definition electrode system-based temporal interfering stimulation (HD-tTIS)~\cite{huang2020optimization}.
With these advances in tES, tES-based neuromodulation has been reported effective in many applications, such as memory consolidation~\cite{lustenberger2016feedback}, learning enhancement ~\cite{wischnewski2016effects} and treatment of brain disease ~\cite{Meinzer2016ElectricalSO,Inukai2016ComparisonOT}. However, the same neuromodulation strategy may produce various outcomes on a group of participants, demonstrating inter-subject variability~\cite{sadeghihassanabadi2022optimizing}.
This inter-subject variability should be considered rather than neglected when designing personalized stimulation strategies. To this end, computational models are needed to guide personalized tES stimulation strategy, to clarify how the inter-subject variability affects the performance of the stimulation strategy, and to improve the effectiveness of tES. 

A variety of methods have been proposed for optimizing electrical stimulation strategy in tACS and tDCS ~\cite{huang2018optimized,caulfield2020transcranial,dmochowski2011optimized,guler2016optimization,ruffini2014optimization,caulfield2021optimizing,sadeghihassanabadi2022optimizing,saturnino2021optimizing,sadleir2012target}. Unfortunately, these methods cannot be directly applied to tTIS optimization due to the nonlinear and non-convex nature of the tTIS optimization problem.
To solve this non-convex problem, some methods for tTIS policy optimization have been proposed~\cite{lee2020individually,rampersad2019prospects,huang2020optimization,stoupis2022non,huang2019can}. 
First, the exhaustive search algorithm is proposed, by traversing all possible solutions. Its advantage is that it can optimize the electrode montage and the corresponding injection current, and obtain the unique, optimal solution based on the whole-brain electric field~\cite{lee2020individually, rampersad2019prospects}; however, it is too time-consuming, especially for personalized clinical applications.
Second, non-convex optimization methods, such as successive convex approximation, are used to approximate the optimal solution. They formulate objective functions and constraints, such as the weighted least squares with constraints or the minimum variance with linear constraints, and construct approximation functions to solve the optimization problem~\cite{nie2011robust}. These approximation methods are widely used for TES optimization, including mainstream optimization software such as SimNIBS~\cite{saturnino2019simnibs,saturnino2021optimizing} and Roast~\cite{huang2019realistic}, as well as for tTIS optimization~\cite{huang2020optimization,huang2019can}. However, these methods have poor scalability and are difficult to optimize under some complex constraints. For instance, non-convex optimization needs to be combined with a branch and bound algorithm to solve optimization with limited electrical sources~\cite{saturnino2019accessibility}.
In addition, machine learning has been used to optimize electrode montages. For example, Ruffini et al. employed the genetic algorithm to optimize multi-focal stimulation~\cite{ruffini2014optimization}, and Stoupis et al. used the genetic algorithm to obtain a focal stimulation over the hippocampus and thalamus using two pairs of tTIS electrodes~\cite{stoupis2022non}. Bahn et al. applied unsupervised neural networks to optimize TES and realize multi-target stimulation based on HD-tTIS and HD-tDCS~\cite{bahn2021computational}.
However, the above machine learning methods are all designed for a single objective function or the ratio of multiple objective functions. They can only generate a single result, which makes it challenging to obtain a specific strength result and analyze the trade-offs between mutually exclusive conditions. Furthermore, although some methods can be run multiple times to obtain trade-offs by adjusting weights or thresholds, the reasonable range of parameters still need to be set manually~\cite{dmochowski2011optimized, huang2020optimization}.

To obtain trade-offs between mutually exclusive conditions, we introduce the concept of a constrained multi-objective optimization problem (CMOP) and an evolutionary algorithm for solving CMOP. CMOPs are typically non-convex problems due to the discontinuity of objective functions, and potential conflicts between objectives in CMOP make it difficult for traditional algorithms to obtain globally optimal solutions. Therefore, the trade-offs between competing objectives in CMOP generate the Pareto front, which is a set of optimal solutions. Evolutionary algorithms are well-known effective tools for solving these multi-objective problems \cite{Deb2001MultiobjectiveOU}. They can be combined with Pareto relations, eliminate inferior solutions iteratively, and obtain optimal solutions. These solutions can be projected onto Pareto fronts for explicit visual comparisons of optimal solutions. This combination has been widely used in resource allocation problems, such as energy-efficient shop scheduling \cite{Dai2019MultiobjectiveOF} and task offloading in edge computation \cite{Bozorgchenani2021MultiObjectiveCS}. However, more investigations are needed to define objective functions in CMOP and empower evolutionary algorithms with appropriate optimizing search rules.

Here, we present a multi-objective optimization via evolutionary algorithm (MOVEA) for searching optimal tES policy. 
The MOVEA algorithm provides a general framework to solve CMOP for a variety of tES stimulation technology, including tACS, tTIS and HD-tTIS. This framework allows flexibly addition of objective functions and constraints, such as defining the direction of maximum field strength and preventing activation of avoidance regions. The major contributions of this work are summarized as follows.
\begin{itemize}
    \item MOVEA defines an effective optimizing search rule and offers a comprehensive comparison of various stimulation methods, revealing that HD-tACS and HD-tTIS have equivalent maximum intensity, while the two-pair  tTIS has a weaker intensity in comparison. (Study 1 \& Study 2)
    \item MOVEA's flexibility in designing objectives and constraints, such as focal stimulation and avoidance areas, demonstrates that HD-tTIS exhibits superior steerability and focality. (Study 1 \& Study 3)
    \item MOVEA highlights the impact of inter-subject variability on tES performance, emphasizing the need for personalized stimulation protocols. (Study 4)
\end{itemize}

\section{Method}

\begin{figure}[thbp]  
	\centering
	\includegraphics[width=0.98\linewidth]{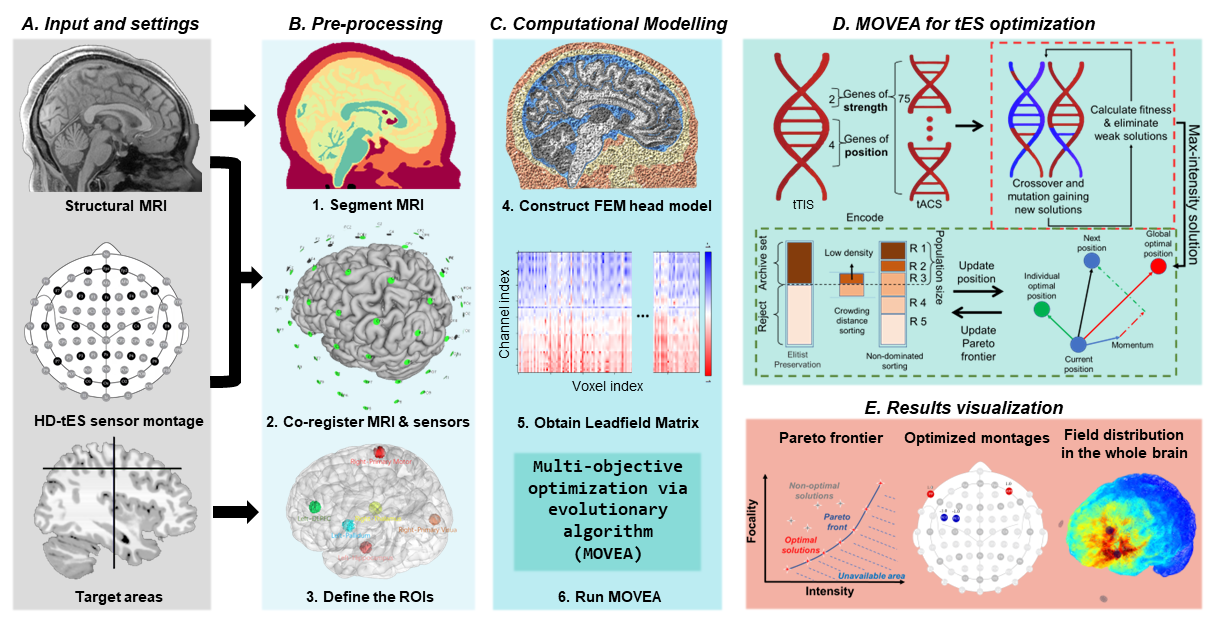}
	\caption{The framework of MOVEA for tES optimization. 
	(A) \textit{Input and setting}, consisting of individual T1-weighted or T2-weighted MR images (top), the HD-tES sensor montage (middle), and the selected target area or avoiding area (bottom).
	(B) \textit{Pre-processing}, to segment the structural MRI (step 1), to co-register the brain with HD-tES sensors for aligning electrodes to the target space (step 2), and then define the ROIs by a sphere with a radius of $10$ $mm$ centred at the target area (step 3).
    (C) \textit{Computational modelling}, to construct a  realistic head model using the finite element method based on Laplace’s equation and electrical conductivity of each brain tissue (step 4), to obtain the leadfield matrix (LFM), which contains information on the distribution of electric fields in each voxel when a current is injected by each electrode (step 5), to run MOVEA with LFM and ROIs as input (step 6).
    (D) \textit{Pipeline of MOVEA for tES optimization}. First, the activated channels and the corresponding current intensities are encoded to sites on the gene. Second, the solution for the electric field intensity at the target ROI is calculated by crossover and mutation (Red box), which is then set as the initial value for the multi-objective optimization problem. For multi-objective optimization, in the decision space, the next position of each solution is determined by its own velocity, its historical optimal solution, and the globally optimal solution (Right of the green box). In the objective space, the dominant solution set is obtained by non-dominated sorting, crowding distance sorting, and elitist preservation (Left of the green box). After several iterations, the Pareto front is obtained.
    (E) \textit{Results visualization}, to illustrate and quantify the results of MOVEA, which includes the Pareto front for target intensity versus focality (left), the optimized tES policy (\ie locations of tES electrodes \& injected currents, middle), the whole-brain field distribution induced by tES (right).}
	\label{fig:framework}
\end{figure}

In this section, we present a customized framework for multi-objective optimization of tES (Fig.~\ref{fig:framework}) 
which can maximize the module of the electric field without imposing a predefined direction. Our method requires individual structural MRI, tES sensor montage and target areas as inputs. First, we pre-process these inputs: segmenting brain volume into tissues with different conductivity, co-registering individual structural MRI with HD-tES sensors, and defining target areas based on Brodmann atlas~\cite{lacadie2008brodmann}. Then based on these pre-processed data, we can construct an accurate finite element method (FEM) head model and calculate the leadfield matrix for the selected ROI. This leadfield matrix is further utilized in the multi-objective optimization via evolutionary algorithm (MOVEA) for tES optimization. The detailed pipeline of MOVEA is illustrated in Fig.~\ref{fig:framework}D. MOVEA provides a Pareto front that contains a set of optimal montages with trade-offs between conflicting objectives as well as the corresponding whole-brain electrical distributions produced by these montages.
    
\subsection{Construction of realistic head model} \label{settings}

The head volume conductor model, also known as the forward model or leadfield matrix in the tES community, contains the mapping relationship between tES stimulation over the scalp and the induced electrical field in the brain. A variety of methods have been developed to construct a realistic head model, including the boundary element method and the finite element method (FEM). In this study, we choose to employ the FEM approach due to its superior accuracy and computational efficiency \cite{saturnino2019electric}. We calculate the leadfield matrix based on individual structural MRI and HD-tES sensor montage. Specifically, we first segment the head tissues from individual T1-weighted and T2-weighted MR images and then create the FEM head model by \textit{Headreco} toolbox ~\cite{nielsen2018automatic}.
Our head model focuses on simulating the effects of extrinsic electrical stimulation. Since the electrical signals within the brain are too  weak to be considered in this problem, the model can be simplified to have no sources or sinks, where the Laplace equation (Eq.~\eqref{eq:Laplace}) can be used to calculate the distribution $ V $ in the volume~\cite{griffiths1999college}. 

\begin{equation}
\label{eq:Laplace}
\nabla \cdot \left( {\sigma \nabla V} \right) = 0,
\end{equation}
where $\sigma$ is the conductivity of tissue which is assumed to be isotropic. The conductivity of tissues are assigned to \textit{0.126} $S/m$ (white matter), \textit{0.276} $S/m$ (gray matter), \textit{1.65} $S/m$ (cerebrospinal fluid (CSF) ), \textit{0.01} $S/m$ (skull), \textit{0.465} $S/m$ (scalp), \textit{0.5} $S/m$ (eye), \textit{2.5} $e^{-14} S/m$ (air cavities), 
\textit{1.0} $S/m$ (saline), according to previous literatures~\cite{1325819,saturnino2015importance,opitz2015determinants}.

\subsection{Multi-objective optimization via evolutionary algorithm (MOVEA)}

We consider this mission to be a CMOP and its definition with $M$ objective functions is as follows~\eqref{eq:objectives}. In this study, the commonly use objective functions are \textit{intensity} and \textit{focality}, which refers to the averaged electric field strength across finite elements within ROI and the whole brain, respectively. The goal is to achieve maximum intensity and minimum focal length,  subjecting to safety constraints, as shown in Eq.~\eqref{eq:objectives}. 

\begin{align}
\label{eq:objectives}
\begin{array}{cc}
&\min/\max \quad f_m(x), \: m = 1, 2, ..., M \\
& s.t. \quad \text{safety constraints},
\end{array}
\end{align}
where $x$ denotes the index of the activation electrode and the corresponding current intensity, and $f_m(x)$ represents the $m$-th objective function ($m \in \{1, 2, ..., M\}$). 
We will introduce the MOVEA algorithm to solve this CMOP problem in subsequent following subsections.

\subsubsection{Simulating the tACS-induced electric field in the brain}

Considering the input current $ s $ 

applied by tES electrodes excluding reference electrode), the induced electric field in the brain can be computed with the leadfield matrix $A$ as $ E = As $, where $ E $ denotes the triaxial electric field intensity. According to \cite{saturnino2019accessibility}, the averaged electric field intensity across a brain area $ \Omega $ in the field direction, $E_\Omega$, can be formulated as Eq.~\eqref{eq:tACS-intensity}.

\begin{equation} 
E_\Omega = \frac{1}{G_\Omega} \int_{\Omega} \bm{{\Vert E \Vert}}{d}G = \frac{1}{G_\Omega} \int_{\Omega} \sqrt{w_xE_x^{2}+w_yE_y^{2}+w_zE_z^{2}}{d}G,
\label{eq:tACS-intensity}
\end{equation}
where $G_\Omega$ denotes the total volume in the brain area $\Omega$. $x$, $y$ and $z$ represent three orthogonal directions, and $w_x$, $w_y$, and $w_z$ denote the weights assigned to the respective orthogonal directions.  In this study, the objective is to maximize the module of the electric field for target area $E_\Omega$ rather than the directional electric field. As such, the values of $w_x$, $w_y$, and $w_z$ are all equivalent to unity.

\subsubsection{Simulating the TI-induced electric field in the brain}

In this study, we investigate two types of tTIS, specifically two-pair tTIS and HD-tTIS. Two-pair tTIS utilizes two pairs of electrodes, while HD-tTIS employs an array of electrodes for stimulation \cite{huang2020optimization}. Both types of stimulation are designed to generate two distinct frequency electric fields that interact to produce a modulated envelope field within the brain. As we disregard the possible effects of high-frequency stimulation on neurons, we calculate the distinct frequency electric field in the same manner as tACS (Eq.~\eqref{eq:tACS-intensity}).

The envelope field with desired direction can be calculated by the formula proposed by Grossman et al.~\cite{grossman2017noninvasive}, shown in Eq.~\ref{eq:ti2}.
\begin{equation}
|\vec{E}(\vec{r})|= || (\vec{E}_{1}(\vec{r}) + \vec{E}_{2}(\vec{r})) \cdot (\vec{n}) |
- | (\vec{E}_{1}(\vec{r}) - \vec{E}_{2}(\vec{r})) \cdot (\vec{n}) ||,
\label{eq:ti2}
\end{equation}
where $\vec{E}_{1}(\vec{r})$ and $\vec{E}_{2}(\vec{r})$ are the first and second distinct fields at the location $\vec{r}$ and $\vec{n}$ is an unit vector along the desired direction.

The envelope field in the brain without a predefined direction can be calculated as Eq.~\eqref{eq:ti}:

\begin{equation}
\left|\vec{E}_{AM}(\vec{r})\right|=\left\{\begin{array}{l}
 2\left|\vec{E}_{2}(\vec{r})\right|,   \qquad\qquad\qquad\qquad \qquad\qquad\qquad\qquad \: |\vec{E}_{1}(\vec{r})| < |\vec{E}_{2}(\vec{r})| cos(\alpha)\\
2\left|\vec{E}_{2}(\vec{r}) \times\left(\vec{E}_{1}(\vec{r})-\vec{E}_{2}(\vec{r})\right)\right| /\left|\vec{E}_{1}(\vec{r})-\vec{E}_{2}(\vec{r})\right|, \: otherwise
\end{array}\right.
\label{eq:ti}
\end{equation}

where $\alpha$ is the angle between $\vec{E}_{1} $ and $ \vec{E}_{2} $. The Eq.~\eqref{eq:ti} holds only when $\alpha$ < 90° and $|\vec{E}_{1}(\vec{r})| > |\vec{E}_{2}(\vec{r})|$, or the sign of $ \vec{E}_{2} $ and the numbering of the channels should be inverted and swapped respectively.

\subsubsection{Designing the safety constraints}\label{constraints}

There are some additional constraints to ensure the safety of human participants. The total injection current and the maximum individual injection current are denoted as $ I_{tot} $ and $ I_{ind} $, respectively. For $N$ candidate electrodes, considering the presence of a reference electrode and according to Kirchhoff’s current law, Eq.~\eqref{con1} constrains that the sum of the absolute values of the current intensity of candidates and reference electrode should be smaller than $ 2 * I_{tot} $. The constraints, Eq.~\eqref{con2} and Eq.~\eqref{con3} are set to avoid skin irritation, discomfort, and heating at the individual electrode interface.

\begin{equation}
{g_1}\left( s \right) = \sum\nolimits_n {\left| {{s_n}} \right|} + \left| {\sum\nolimits_n {{s_n}} } \right| \leq 2{I_{tot}}, \label{con1}
\end{equation}

\begin{equation}
{g_2}(s) = \left| {{s_n}} \right| \leq {I_{ind}}, \label{con2}
\end{equation}

\begin{equation}
{g_3}\left( s \right) = \left| {\sum\nolimits_n {{s_n}} } \right| \leq {I_{ind}}, \label{con3}
\end{equation}

To quantify the degree of violation to constraints~\cite{woldesenbet2009constraint}, we define a constraint violation ($CV$) index in Eq.\eqref{con4}. Since the constraint Eq.\eqref{con2} is guaranteed during the encoding (Fig.\ref{fig:framework}D), $CV$ is a summation of the normalized violations to the constraint Eq. (\ref{con1} \& \ref{con3}).

\begin{equation}
CV = max(\frac{{\left| {\sum\nolimits_n {{s_n}} }\right| - {I_{ind}}}}{I_{ind}},0) + max(\frac{{ \sum\nolimits_n {\left| {{s_n}} \right|} + \left| {\sum\nolimits_n {{s_n}} } \right| - {2 I_{tot}} }}{2{I_{tot}}},0),
\label{con4}
\end{equation}
where the function $max(x,0)$ outputs the maximal value between $x$ and 0.
In order to ensure a fair comparison, a consistent total injected current of 2 $mA$ was maintained for each stimulation in this study. Subsequently, individual injected currents were applied based on the type of stimulation used, with the HD electrode system set at 1 $mA$ \cite{dmochowski2011optimized,rampersad2019prospects,ruffini2014optimization}and the two-pair tTIS set at 1.5 $mA$\cite{stoupis2022non,lee2020individually}. Additionally, when using two-pair  tTIS , there is an extra constraint that must be considered to prevent the selection of the same channel by both pairs of electrodes.

\subsubsection{Formulating CMOP for optimizing tES policy}\label{pareto_sort}

Realistic electrical stimulation tasks often require multiple goals to be met in addition to safety limitations. 
To address this, we formulate a CMOP problem for optimizing tES policy. 
Using focal stimulation as an example, our optimization aims to achieve the strongest electric field \textit{intensity} in the target region with the highest \textit{focality} (\ie the smallest whole-brain electric field intensity). A solution involves the identification of the optimal combination of stimulation electrodes within the standard 10-10 system, as well as the determination of the injected current for each electrode.

In practice, it is often challenging to simultaneously optimize the intensity and focality of transcranial electrical stimulation. It is difficult to have a solution that achieves maximum intensity and focality at the same time. Consequently, the outcome of the multi-objective optimization problem is a set of optimal solutions represented by a \textit{Pareto front}. This Pareto front comprises solutions that optimize multiple objective functions, such as intensity and focality. The performance of these solutions on any given objective function cannot be further improved without a corresponding degradation in the performance of other objective functions. Existing methods often obtain a Pareto front by changing thresholds or weights and running it multiple times\cite{dmochowski2011optimized,huang2020optimization}. Our algorithm utilizes Pareto optimization to generate the Pareto front instead of adjusting parameters, allowing for computationally efficient control of the search direction. This is particularly beneficial when facing more than two objections, as it enables the effective handling of complex optimization problems.

We used Pareto optimization for the comparison of solutions and generate the Pareto front. For $M$ minimized targets, we say the solution $s_a$ \textit{Pareto dominates} the solution $s_b$ if the solution $s_a$ does not perform worse than the solution $s_b$ on any objective function and the solution $s_a$ performs better than the solution $s_b$ on some objective functions. ($s_b \prec s_a \Leftrightarrow  \{\forall i \in \{1,...,M\} : f_i(s_b) \geq f_i(s_a)\} \land  \{ \exists j \in \{1,...,M\} : f_j(s_b) > f_j(s_a)\} $ ) Thus, the Pareto front contains all solutions that are not Pareto dominated by others.

In cases where constraint violations are considered, the Pareto front is more complicated since the constraints in the multi-objective optimization problem may be satisfied or violated. 
To tackle this issue, we prioritize minimizing $CV$ index and incorporate it into the Pareto dominance criterion. This additional setting directly rules out unsatisfied constraints and narrows down the search space in the optimization process, resulting in solutions that meet all requirements. 

We say the individual \textit{a} Pareto dominates the individual \textit{b} if one of the following conditions are met: 
\begin{itemize}
    \item Individual \textit{a} satisfies the constraint, but individual \textit{b} violates it; 
    \item Both the individual \textit{a} and \textit{b} meet the constraint conditions, and individual \textit{a} Pareto dominates individual \textit{b} in all other indicators;
    \item Both the individual \textit{a} and individual \textit{b} are not satisfied, but the constraint violation of individual \textit{a} is smaller.
\end{itemize}
In the implementation of MOVEA, we utilize a penalty function to manage the priority of solutions and prevent getting trapped in solutions with an undesirably small field (Algorithm $Evaluate_solution$). Once the priority relations have been addressed, the fast nondominated sorting technique is employed to expedite the ranking process~\cite{chaube2012adapted}.
 
As the optimization process is based on Pareto optimization, scaling to situations with more objectives becomes more flexible and straightforward. By incorporating avoidance zones into the focal optimization process, we enable the simultaneous optimization of three objectives. Notably, the Pareto front in this case will expand from a curve to a surface, effectively adapting to the increased complexity of the multi-objective optimization problem.

\subsubsection{Solving CMOP via evolutionary algorithm}

In this study, we employ the evolutionary algorithm to solve the CMOP. 
However, safety constraints can lead to multi-objective algorithms such as NSGAII~\cite{chaube2012adapted} yielding infeasible solutions in the initial epochs when directly initializing. 

Additionally, the nonlinearity and non-convexity of the tES model require an extended duration for the multi-objective algorithm to identify optimal electrode montages with the highest electric field intensity. Given the intricacies involved in multi-objective optimization, enhancing algorithms' efficiency to accelerate the optimization process is crucial.

We propose a hybrid approach combining genetic algorithms and particle swarm algorithms (Fig.~\ref{fig:framework}D). Specifically, we first use a single-objective genetic algorithm to find the maximum intensity target and then apply it as prior knowledge to search the Pareto front using multiple objective particle swarm optimization (MOPSO) \cite{coello2002mopso}. Pseudocode of Algorithm can be found in supplementary.

In MOVEA, we begin by encoding the montages to make them readable by the algorithm and ensure that safety constraint (Eq.~\ref{con2}) is met in the process (Fig.~\ref{fig:framework} D). For tACS, we use an array with a length of $N$, each point representing the injected current strength of the corresponding channel. For tTIS, we use an array with a length of $6$. The first two values represent the strength of each pair of electrodes, and the last four represent the index of the selected channel. For HD-tTIS, we utilize an array of length $2N$ for this approach. The array is partitioned into two segments, representing the injected current strength for each of the two HD-electrode systems. 
We then computed the envelope using the electrical field generated by these two high-density  electrode systems derived from Eq.(\ref{eq:ti}).

In the first stage, we employ a genetic algorithm for the single-objective search as depicted in Algorithm $Stage_{One}$.To obtain high intensity in a short period of time, we set the injected energy to the maximum value within the safety constraints, limiting the current intensity to 1 or -1 $mA$. This design facilitates the algorithm to circumvent safety constraints during the first stage, thereby enabling faster convergence to the optimal solution. Inspired by biological genetics, 
we consider the encoded array as a gene, and derive  new phenotypes through operations such as mutation and crossover~\cite{hassanat2019choosing}. We compare the new solutions with previous solutions, eliminating inferior solutions. After several iterations, we obtain the solution with the maximum intensity.

In the second stage, we implement MOPSO to obtain the Pareto front of trade-off targets (Algorithm $Stage_Two$). The MOPSO considers a solution as a particle, where the values in the array correspond to the coordinates of each dimension in a high-dimensional decision space. And each particle possesses a velocity variable that indicates its rate of evolution. As a heuristic algorithm, MOPSO maintains previously discovered non-dominated particles as the $archive$ and leverages them to influence the evolution of other particles.
At first, one particle's position is initialized using the prior knowledge obtained in $Stage\_One$, while the other particles' positions are initialized according to Algorithm $Initiation$ to ensure diversity. If all particle positions are initialized with the prior knowledge from $Stage One$, the population may become trapped in that particular position. During initiation, the $Sparse constant$ in Algorithm $Initiation$ is employed to balance between activating too many channels, which could violate safety constraints, or activating insufficient channels, potentially resulting in an ineffective field. And the objective here is to match the expected injected power with the safety constraint values. Thus, a Monte Carlo simulation is employed to compute the expected of injected power as outlined in Algorithm $Initiation$ and Eq.~\eqref{con1}. The Monte Carlo simulation is implemented with Scipy library~\cite{2020SciPy-NMeth}. When $I_{(tot)}$ is equal to 2, the $Sparse \thinspace constant$ is set to 5 for HD-tACS (75 channels) and 9 for HD-tTIS (150 channels). Furthermore, the velocity of the particle, denoting the search speed, is randomly initialized between 0 and 1.

During this stage, we account for all objectives and relax the constraints based on the aforementioned assumption enabling the particles to explore the Pareto front more flexibly. Consequently, we evaluate the particles using the Pareto optimization, as previously described, and add the non-dominated particles to the $archive$.
If the size of the $archive$ surpasses the limit, we adopt the $Crowding Distance Sorting$ approach (Algorithm $Crowding\_Distance\_Sorting$) to eliminate the excess points, which is illustrated in the bottom left corner of Fig.~\ref{fig:framework} D.
Particles calculate the subsequent position according to their optimal historical position and the optimal global position from the $archive$, as the following formula. 

\begin{equation}
v_i(t+1) = w_1 * v_i(t) + c_1 * rand() * (pbest_i - x_i(t)) +  c_2 * rand() * (gbest - p_i(t)) \label{eq:velocity-update}
\end{equation}

\begin{equation}
p_i(t+1) = p_i(t) + v_i(t+1), \label{eq:position-update}
\end{equation}
where $v_i(t)$ is the velocity of particle $i$ at iteration $t$, $p_i(t)$ is its position, $pbest_i$ is the best position found by particle $i$ so far, $gbest$ is the global best position chose from $archive$ randomly , $w$, $c_1$, and $c_2$ are constants to control the influence of the current velocity, personal best position, and global best position, respectively, their values are set to 0.4, 2, and 2 in this study.  The $rand()$function samples a number from a uniform distribution over the interval $[0 ,1]$.

During the early stages of evolution, the particle with prior knowledge represents the optimal solution and is consequently added to the $archive$. This enables other particles to rapidly congregate around it in the parameter space, leading to the emergence of Pareto front. Once we obtain the new positions, we evaluate the particles' performance corresponding to their current position and determine whether to update their optimal position.

For each iteration, we select the superior solutions based the Pareto relationships, and if they do not Pareto dominate each other, we choose the solutions for particles located in less populated areas in the objective space, using Algorithm $Crowding\_Distance\_Sorting$ to ensure the even distribution of Pareto front.

\section{Experiments and Results}

\subsection{Dataset and experimental setup}
\label{sec:setting}

In this work, we use the T1 and T2 MRI data from the HCP S1200 Release 
dataset~\cite{van2013wu}.
The FEM head model and the associated comparison algorithm were implemented using SimNIBS 3.2.6\cite{saturnino2019simnibs}.
The parameters were set to their default values, and the desired direction was set to $None$. The MOVEA algorithm was implemented using Python 3.8, with a population size of 30 chromosomes and a termination criterion of 50 iterations for $Stage\_One$, and a population and archive size of 100 with a termination criterion of 100 iterations for $Stage\_Two$. All statistical analyses for the electric field distribution of brain voxels are conducted with $Origin$ (Version 2022. OriginLab Corporation, Northampton, MA, USA).

\subsection{Study 1: Optimization of tES for six target regions.}
In order to compare the effects of optimized tES, we perform the MOVEA algorithm on the same head model to achieve Pareto fronts for different brain regions. We select six target regions of interest (ROIs), three of which are located in superficial areas of the brain (i.e., motor cortex, dorsolateral prefrontal cortex [DLPFC], and primary visual cortex [V1]), and three in deep brain areas (i.e., hippocampus, pallidum, and thalamus). These regions have been commonly targeted in previous tES studies \cite{cao2019stimulus, huang2020optimization, huang2021comparison, stoupis2022non}.

Here are some basic settings for this experiment. We define the target region as a 10 $mm$ sphere at the target point. The target point is selected based on the Brodmann atlas, and the value is fine-tuned to emphasize the differences in electric field intensity at varying depths. The MOVEA algorithm aims to maximize the electric field intensity inside the target region while minimizing the average electric field across the brain. The orientation of the electric field is free. The safety constraints are described in detail in Sec.~\ref{constraints}. It is worth noting that that the $I_{ind}$ for tTIS is 1.5 $V/m$, rather than 1 $V/m$\cite{stoupis2022non,lee2020individually}.

\begin{figure}[thbp]  
	\centering
	\includegraphics[width=0.98\linewidth]{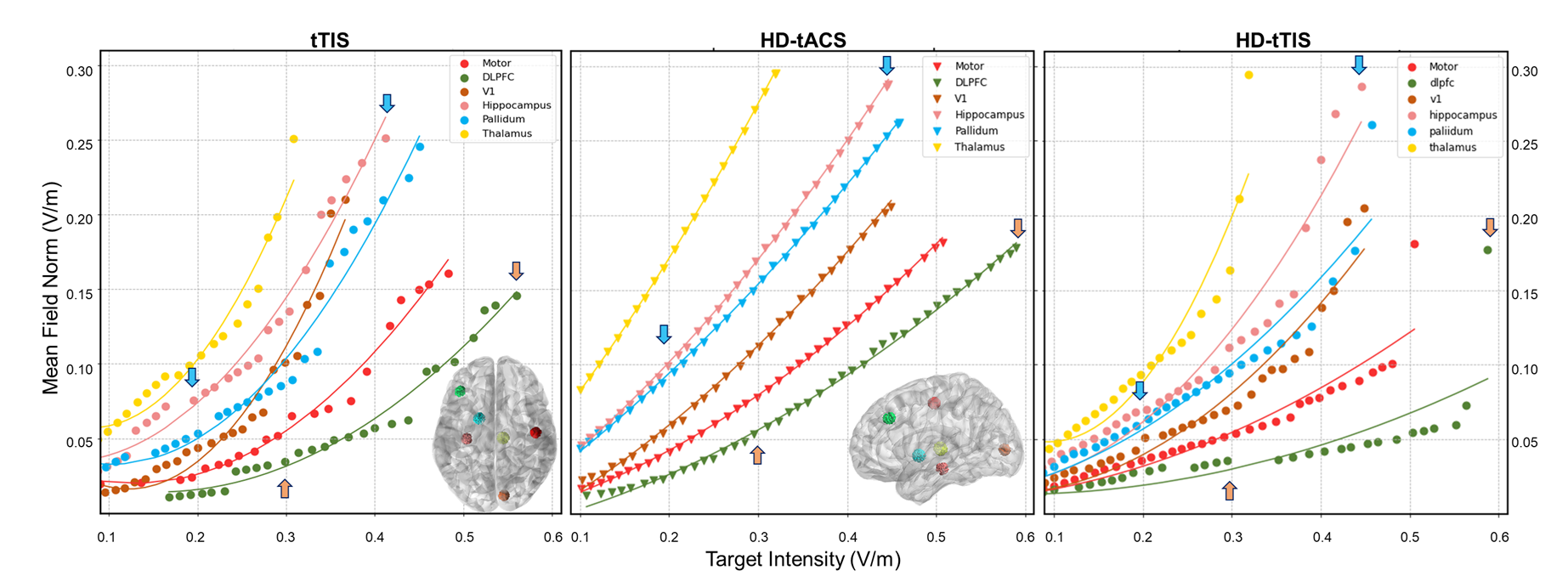}
	\caption{The Pareto fronts obtained by MOVEA for two-pair tTIS (left), tACS (middle) and HD-tTIS (right). We set the target area in the motor cortex, DLPFC, V1, hippocampus, pallidum and thalamus. The ROIs with 10 $mm$ radius are shown in the axial and sagittal view. MOVEA optimizes the tES policy for each ROI separately, which generates the Pareto fronts for target intensity versus mean field norm (\ie focality). The Pareto fronts are obtained by using the realistic FEM head model from a specific subject (\ie HCP sub877168). As examples, we visualize the arrow-marked tTIS and tACS policies in Fig.~\ref{fig:visualization}.
    \textit{MNI coordinates}: DLPFC, [-39, 34, 37]; motor, [47, -13, 52];  V1, [14, -99, -3]; hippocampus, [-31, -20, -14]; pallidum, [-17, 3, -1]; thalamus, [10, -19, 6].The line is the result of a second-order fit.}
	\label{fig:paretofrontier}
\end{figure}

We run the intensity optimization experiment based on the above settings, and the results of an example subject (HCP sub877168) are shown here. Fig.~\ref{fig:paretofrontier} illustrates $12$ Pareto fronts for each of the $6$ target ROIs, including three brain areas in the superficial layer (\ie motor cortex, DLPFC, and V1) and three in the deep brain (\ie hippocampus, pallidum, and thalamus).

\begin{figure}[thbp]  
	\centering
	\includegraphics[width=0.98\linewidth]{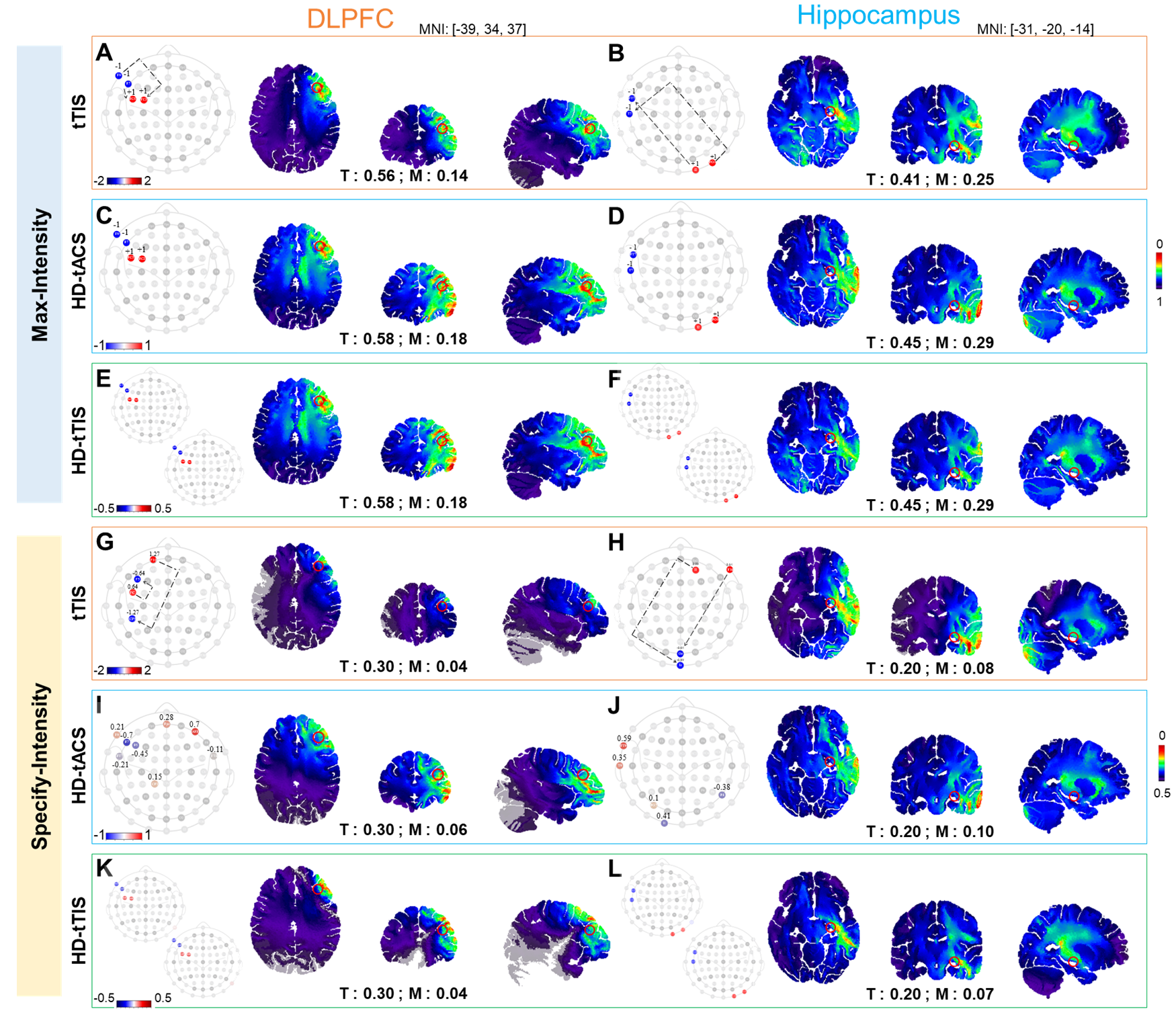}
	\caption{The optimized tES policy by MOVEA for two-pair tTIS (A, D, G, H), HD-tACS (C, D, I, J) and HD-tTIS(E, F, K, L). We present the results for a target area in the \textbf{DLPFC} (A, C, E, G, I, K) and \textbf{hippocampus} (B, D, F, H, J, L), respectively.
	These results correspond to the specific solutions in the Pareto front marked in Fig.~\ref{fig:paretofrontier}. \textbf{Max-Intensity} refers to an objective function to maximize the intensity of the electric field ($T$ in $V/m$ unit) in the target area regardless of the focality ($M$ in $V/m$ unit), while \textbf{Specify-Intensity} refers to an objective function to maximize the focality of the electric field in target area given a fixed electric field intensity $T$ (\ie $T = 0.3$ for DLPFC, $T=0.2$ for hippocampus).  
    In each panel, we illustrate the pairs of activation electrodes and their injected current which is larger than $0.1 mA$ (left), the three views (\ie axial, coronal, and sagittal view) of the induced electric fields in the brain (right), where the red circle indicates the target area. The injected current intensity of each electrode in tES obeys a pre-defined constraint in MOVEA for safety reasons (\ie total injected current: $2mA$ for tTIS and tACS).
    \textit{Abbr.}: tES, transcranial electrical stimulation ;  tTIS, transcranial temporal interference stimulation; HD-tACS, high definition transcranial alternating current stimulation;HD-tTIS, high definition transcranial temporal interference stimulation; DLPFC, dorsolateral prefrontal cortex; $T$, the averaged electric field within the ROI; \textit{M}, the averaged electric field across the whole brain. }
	\label{fig:visualization}
\end{figure}

The Pareto front depicted in Fig.~\ref{fig:paretofrontier} elucidates the inherent trade-off between the intensity and focality of electric field stimulation. Both the choice of the targeted brain region and the stimulation modality play a critical role in determining the optimization outcomes. Notably, when targeting the hippocampus, a deep brain region with a depth of 40 $mm$, the maximum electric field intensity achieved under two-pair tTIS is 0.41 $V/m$, while under tACS and HD-tTIS it improves to 0.45 $V/m$. In contrast, when targeting a more superficial region such as the motor cortex, the maximum intensities under the two stimulation modalities further increase to 0.48 $V/m$ (two-pair tTIS), 0.51 $V/m$ (tACS), and 0.51 $V/m$ (HD-tTIS).

Each point along the Pareto front represents an optimal stimulation strategy, which is a compromise between target intensity and overall focality. Fig.~\ref{fig:visualization} visualizes six example solutions indicated by arrows in Fig.~\ref{fig:paretofrontier}. The visualizations depict the results for two different target regions: the DLPFC (Fig.~\ref{fig:visualization}A, C, E, G, I, K) and the hippocampus (Fig.~\ref{fig:visualization}B, D, F, H, J, L). The figure illustrates the locations of the activation electrodes, their respective injected currents, and the induced electric fields within the brain. For safety considerations, the injected current intensity for each electrode adheres to a predefined constraint (\ie, total injected current 2 $mA$).

In Fig.~\ref{fig:visualization}, we differentiate between two types of solutions: $Max intensity$, which achieves the maximum electric field intensity at the target region, and $Specific intensity$, which achieves a predefined target intensity (0.3 $V/m$ for DLPFC, 0.2 $V/m$ for hippocampus) while maximizing focality. A comparison of these solutions reveals that tES induces electric fields over extensive brain regions in all conditions, leading to a leakage problem where electric fields are present outside the intended target ROIs. The extent of this leakage increases with higher target electric field intensities. However, when the target ROI is located in a superficial brain region, the leakage problem can be mitigated, resulting in a more focal electric field distribution. 

To quantitatively assess and compare the focality of the electric fields generated by two-pair tTIS and tACS, we conduct the MOVEA optimization with a desired target electric field intensity of 0.25 $V/m$. Subsequently, we calculate the proportions of brain voxels exhibiting electric field intensity greater than various threshold levels (\ie, 0.10 $V/m$, 0.20 $V/m$, and 0.25 $V/m$) as shown in Table.~\ref{tab:activate_volume}. The results demonstrate that tTIS can generate more focal electric fields in all six target regions compared to tACS. The table presents the proportions of voxels with electric field intensity greater than the corresponding threshold when the intensity of the target region is 0.25 $V/m$, based on the head model built for HCP subject sub877168. The column $Vol$ represents the proportion of the whole brain volume above the threshold, with smaller values indicating better focality. The column $\delta$ refers to the proportion of reduced activated voxels in tTIS compared to tACS. Notably, as the threshold increases from 0.10 V/m to 0.25 V/m, the thresholded volume ratio for the DLPFC decreases from 6.31\% to 0.38\%. As the most superficial target, the reduced proportion $\delta$ diminishes substantially from 59.81\% to 30.33\%. In contrast, the reduced proportions for other target regions generally increase as the stimulation threshold rises. The volume ratio for tACS in V1 is larger than in the deeper region of the pallidum, indicating a more dispersed electric field in V1. One possible explanation for this observation is the distribution of CSF in the brain. And there is substantial improvement with tTIS stimulation, where $\delta$ increases from 61.33\% to 82.86\%.
Overall, the analysis demonstrates that tTIS is capable of generating more focal electric fields compared to tACS. This capability is important for non-invasive brain stimulation applications where precise targeting of specific brain regions is crucial while minimizing off-target effects.

\begin{table}[thp]

\caption{ The proportion of voxels greater than the corresponding threshold when the intensity of target is $2.5 V/m$. The head model used in MOVEA is built on HCP sub877168. Vol represents the proportion of the whole brain above the threshold. The smaller the Vol, the better the focality. $\Delta$ refers to the proportion of reduced activated voxels compared to $Vol_{tACS}$. }
\renewcommand\arraystretch{1.5}
\begin{center}
\label{tab:activate_volume}
\begin{tabular}{lcccccc}
\hline
\multirow{2}{*}{\textbf{Target}} & \multirow{2}{*}{\textbf{MNI coordinate}} &
\multirow{2}{*}{\textbf{Depth} (mm)} & \multirow{2}{*}{\textbf{Threshold} (V/m)} &
\multicolumn{2}{c}{\textbf{Vol} (\%)} & \multirow{2}{*}{\textbf{$\Delta$} \: (\%)}  \\
\cline{5-6}
& & & & tACS & tTIS & \\ \hline
\multirow{3}{*}{\textbf{DLPFC}} & \multirow{3}{*}{[-39, 34, 37]} & \multirow{3}{*}{7} & 0.10             & 15.7      & \textbf{6.31}     & 59.81       \\
& & & 0.20             & 1.89      & \textbf{0.80}     & 57.7      \\
& & & 0.25            & 0.54      & \textbf{0.38}     & 30.33      \\ \hline
\multirow{3}{*}{\textbf{Motor}} & \multirow{3}{*}{[47, -13, 52]} & \multirow{3}{*}{10} & 0.10             & 26.9      & \textbf{10.7}     & 60.22       \\
& & & 0.20             & 4.32      & \textbf{1.48}     & 65,74       \\
& & & 0.25            & 1.51      & \textbf{0.49}     & 67.15       \\ \hline
\multirow{3}{*}{\textbf{V1}} & \multirow{3}{*}{[14, -99, -3]} & \multirow{3}{*}{20} & 0.10             & 69.3     & \textbf{26.8}     & 61.33      \\
& & & 0.20             & 17.7      & \textbf{3.46}     & 80.45       \\
& & & 0.25            & 8.40      & \textbf{1.44}     & 82.86       \\ \hline
\multirow{3}{*}{\textbf{Pallidum}}  & \multirow{3}{*}{[-17, 3, -1]} & \multirow{3}{*}{45} & 0.10             & 62.6      & \textbf{42.0}     & 32.91       \\
& & & 0.20             & 16.5      & \textbf{6.50}     & 60.30       \\
& & & 0.25            & 6.50      & \textbf{2.29}     & 64.77      \\ \hline
\multirow{3}{*}{\textbf{Hippocampus}} & \multirow{3}{*}{[-31, -20, -14]} & \multirow{3}{*}{40} & 0.10             & 79.8      & \textbf{65.0}     & 18.55      \\
& & & 0.20             & 21.6      & \textbf{11.0}     & 49.07      \\
& & & 0.25            & 8.59      & \textbf{3.27}     & 61.93       \\ \hline
\multirow{3}{*}{\textbf{Thalamus}}  & \multirow{3}{*}{[10, -19, 6]} & \multirow{3}{*}{62} & 0.10             & 99.5      & \textbf{95.5}     & 4.02      \\
& & & 0.20             & 61.5      & \textbf{33.4}     & 45.69       \\
& & & 0.25            & 30.3      & \textbf{14.8}     & 51.16       \\ \hline
\end{tabular}

\end{center}

\end{table}

\subsection{Study 2: The achievable maximal electric field across the whole brain}

The maximum electric field intensity achievable by HD-tACS and HD-tTIS is observed to be the same. To further investigate the achievable maximum intensity by two-pair tTIS and HD-tACS across the whole brain, we conducted experiments to generate the highest intensity at two types of target regions: 1) a group of voxels and 2) a single voxel.

\begin{figure}[htbp]  
	\centering
	\includegraphics[width=1\linewidth]{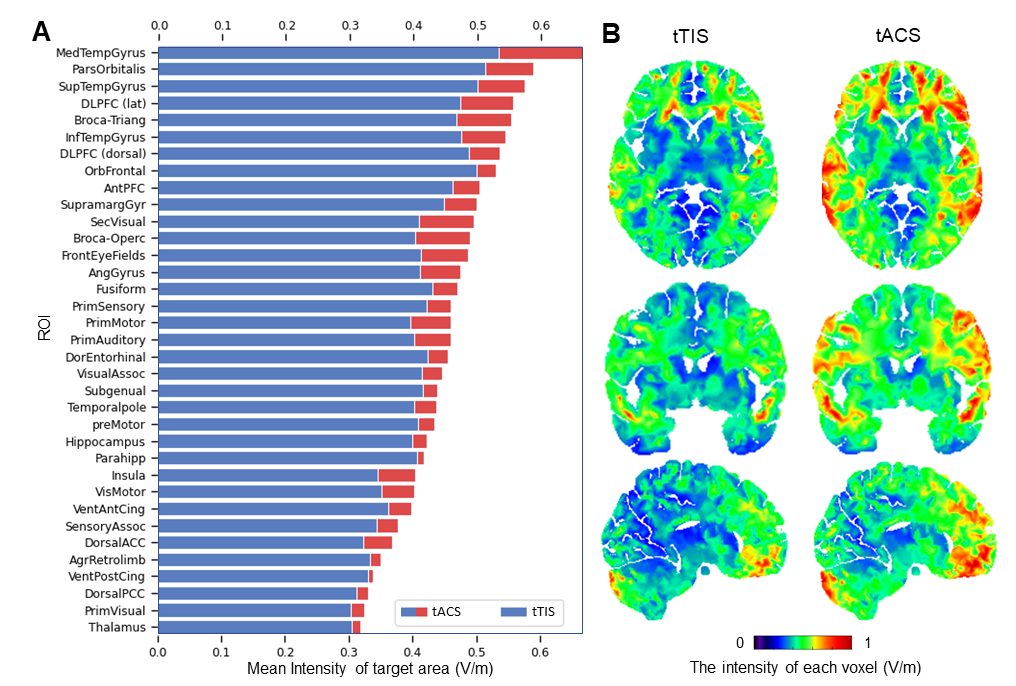}
    \caption{The maximal intensity achieved by MOVEA in each brain region with $2 mA$ of tES on the scalp. (A) The maximal mean intensity achieved in the 35 brain regions. The Brodmann atlas defines the regions. In all 35 brain regions, the current intensity that HD-tACS can achieve is higher than two-pair tTIS, and the increment of HD-tACS over two-pair tTIS is indicated with the red box. (B) The spatial distribution of the maximal intensity achieved by two-pair tTIS (left) and HD-tACS (right). }
	\label{fig:allrois}
\end{figure}

For the first type of target region (group of voxels), we defined a sphere with a radius of 10 $mm$ at the center of each Brodmann region as the target region. We optimized the stimulation policy for two-pair tTIS and HD-tACS using the MOVEA algorithm, imposing a constraint that limits the maximum current injection to 2 $mA$. Fig.~\ref{fig:allrois}A illustrates the maximum electric field intensity achieved in 35 brain regions by two-pair tTIS and HD-tACS. Notably, the achievable intensity is inversely related to the depth of the target region, with deeper regions exhibiting lower maximum intensity.

For the second type of target region (single voxel), we randomly selected individual voxels within the brain and examined the distribution of achievable intensity across the whole brain. We performed 20,000 random samples from a total of 1,467,981 voxels (the total number of voxels in the brain).
The distribution of achievable maximum intensity at the single voxel level is shown in Fig.~\ref{fig:allrois}B. In this setting, HD-tACS demonstrates is observed to achieve higher induced electric field intensities compared to two-pair tTIS. The spatial distribution of maximum intensity induced by HD-tACS is heterogeneous, with stronger electric field intensities being achieved in the superficial regions of the brain. In contrast, the achievable electric field intensity by two-pair tTIS exhibits a more evenly distributed pattern across the brain, as depicted in Fig.~\ref{fig:allrois}B.

\subsection{Study 3: Optimal tES design to avoid activating a specific brain region}

One of the salient advantages of the MOVEA approach is the ability to incorporate a specific brain region, referred to as the avoidance region, into the optimization process to prevent unintended activation. To achieve this, an additional objective representing the electric field intensity in the avoidance region is introduced when calculating the Pareto relationship. In order to evaluate the effectiveness of MOVEA in avoiding stimulation of the avoidance region, we conduct optimization experiments involving the right motor cortex (MNI: [47,-13,52]) as the avoidance region and the right thalamus (MNI: [10,-19,6]) as the target region. The optimal stimulation policies are obtained for the following five experimental setups:
\begin{enumerate}
    \item The tACS setup optimized by SimNIBS, consisting of two conventional $5 cm \times 5 cm$ electrodes, each limited to a maximum 2 $mA$ injection current ($I_{ind}=2$, $I_{tot}=2$); 
    \item The HD-tACS setup optimized by SimNIBS, consisting of any number of electrodes in 10-10 standard system ($I_{ind}=2$, $I_{tot}=1$);
    \item The HD-tACS setup optimized by MOVEA, consisting of any number of electrodes in 10-10 standard system ($I_{ind}=2$, $I_{tot}=1$);
    \item The two-pair tTIS setup optimized by MOVEA, consisting of two pairs of electrodes in 10-10 standard system ($I_{ind}=2$, $I_{tot}=1.5$);
    \item The HD-tTIS setup optimized by MOVEA, consisting of any number of electrodes in 10-10 standard system ($I_{ind}=2$, $I_{tot}=1$);
\end{enumerate}

\begin{figure}[htbp]  
	\centering
	\includegraphics[width=1\linewidth]{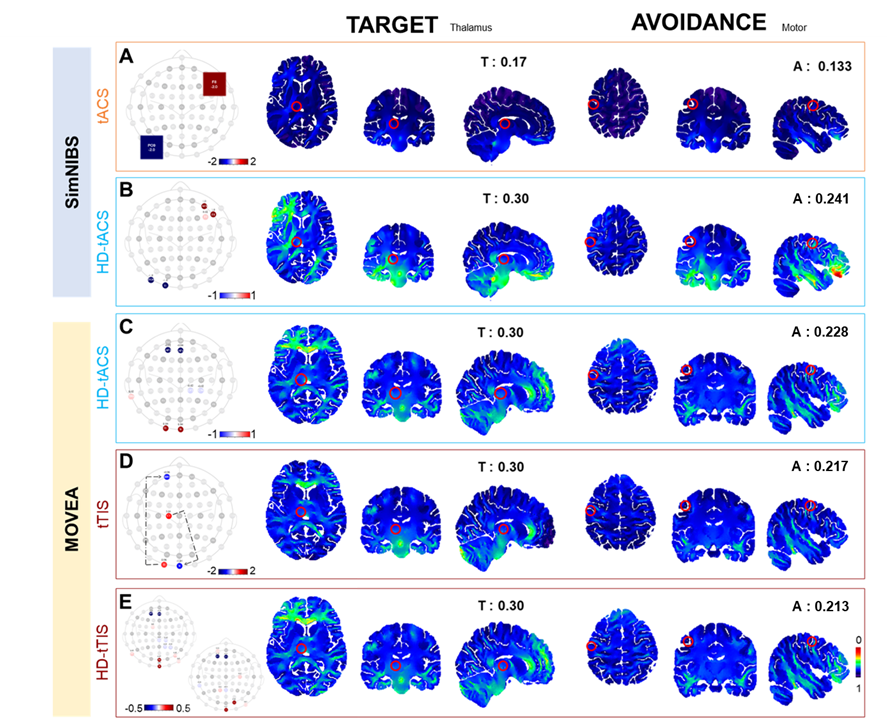}
    \caption{
    Optimization results with avoidance zones. Four stimulation strategies were used.
(A) The conventional electrodes consist of two $5 cm \times 5 cm$ patches, each of which can inject a maximum of 2 $mA$ current;
(B) HD-tACS optimized by SimNIBS;
(C) HD-tACS optimized by MOVEA;
(D) two-pair tTIS optimized by MOVEA;
(E) HD-tTIS optimized by MOVEA. 
The first column of each row represents the electrode distribution, the second column is the visualization of the electric field distribution in the target ROI, and the third column is the visualization of the electric field distribution in the avoidance region.
\textit{Abbr.}: tACS, transcranial alternating current stimulation; HD-tACS, high-Definition  transcranial alternating current stimulation; tTIS, transcranial temporal interference stimulation;HD-tTIS, high-Definition  transcranial temporal interference stimulation; $T$, the averaged electric field within the target ROI; $A$, the averaged electric field within the avoidance zone. The unit of $T$ and $A$ is $V/m$.} 
	\label{fig:study3}.
\end{figure}

The comparisons between SimNIBS and our MOVEA method, as well as between traditional TES and HD-TES, are presented in Fig.\ref{fig:study3}. It is observed that traditional tACS with two electrodes produces weaker electric field stimulation in the target region, with an intensity of $T=0.17V/m$ as depicted in Fig.\ref{fig:study3}A, compared to the HD-tACS approach.
When employing an HD-tACS setup, the stimulation strategies recommended by MOVEA and SimNIBS differ. Specifically, MOVEA is able to achieve a lower electric field strength in the avoidance zone ($A=0.228V/m$), as depicted in Fig.\ref{fig:study3}C, in comparison to the electric field strength achieved by SimNIBS ($A=0.241V/m$) shown in Fig.\ref{fig:study3}B.
Given a target electric field intensity of $T=0.3V/m$ in the target area, the high-definition electrode montage has a minimal contribution to the avoidance area. As a result, electric field strengths of $A=0.217V/m$ for two pairs of tTIS (Fig.\ref{fig:study3}D) and 
$A=0.213V/m$ for HD-tTIS (Fig.\ref{fig:study3}E) are achieved.

In summary, tTIS demonstrates greater steerability compared to tACS, particularly when considering the enhancement of electric field intensity in the target region while simultaneously suppressing stimulation in the avoidance region. This advantage makes tTIS a preferred choice for precise and targeted non-invasive brain stimulation interventions.

\subsection{Study 4: Inter-subject variability}

Individual head models exhibit a high degree of inter-subject variability due to differences in head shape and spatial tissue distribution. To investigate the impact of this variability on the optimization of tES protocols, we select eight subjects from HCP dataset~\cite{van2013wu} and construct individual head models based on their T1-weighted and T2-weighted MRI images.

\begin{figure}[h]
	\centering
	\includegraphics[width=0.98\linewidth]{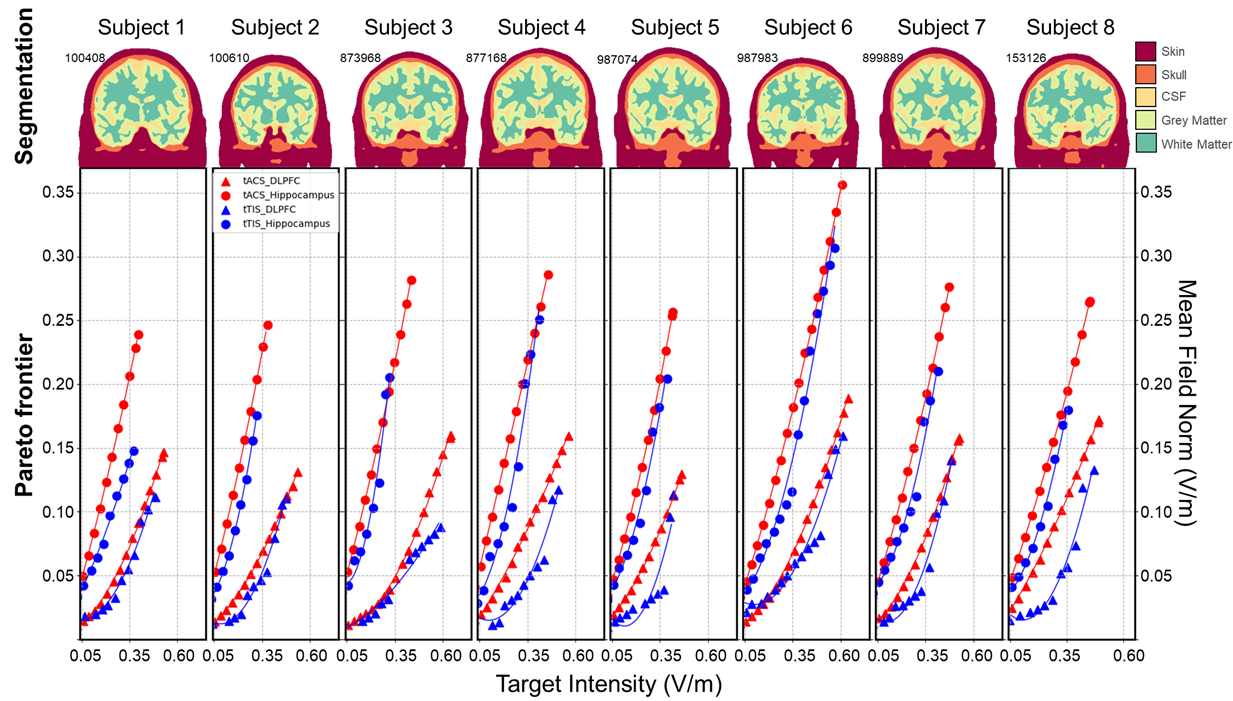}
	\caption{Sensitivity to the personalized head model. The results are from 8 subjects in HCP dataset, showing inter-subject variability. (Up) Segmented tissues from individual T1 MRI, including skin, skull, CSF, grey matter, and white matter. The number shows the index of the HCP subject.  (Bottom) Pareto front optimized with the target area in the DLPFC (triangles) or in the hippocampus (circles). The results of tTIS and tACS are indicated in red and blue, respectively. Each point in the Pareto fronts is an optimal tES strategy with a specific target intensity.The line is the result of a second-order fit.
	}
	\label{fig:pareot_results}
\end{figure}

\begin{figure}[h]
	\centering
	\includegraphics[width=0.98\linewidth]{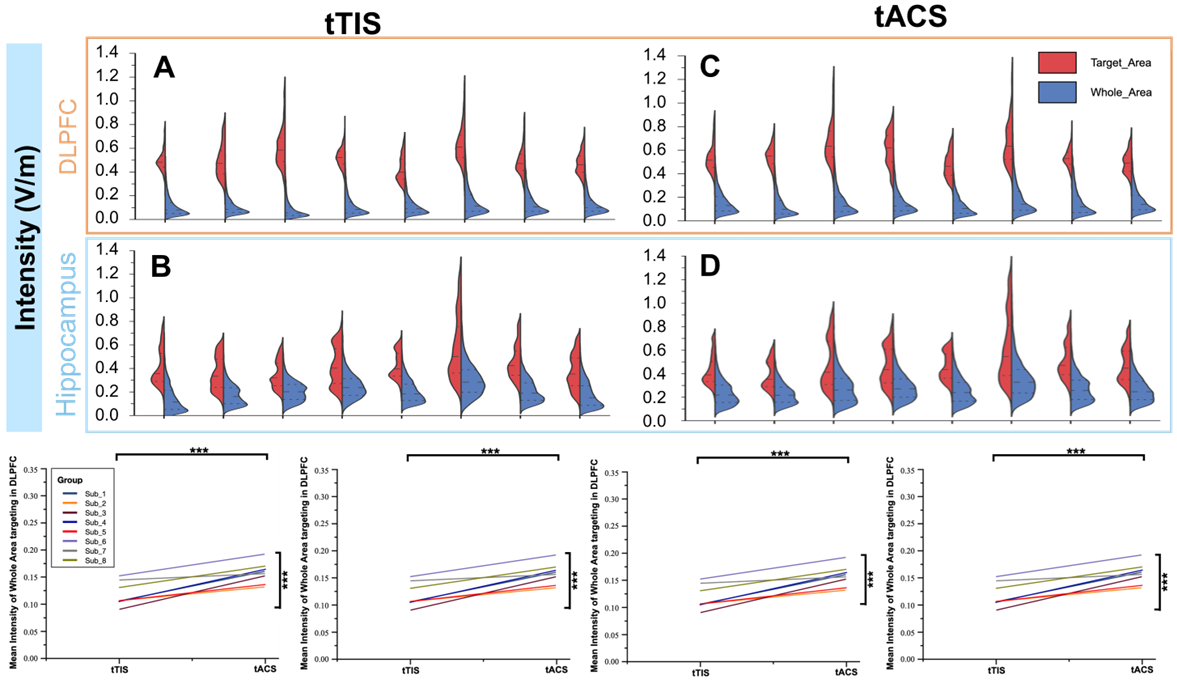}
	\caption{Statistical analysis of electric field strength distribution in eight subjects. We calculated the proportion of voxels at each electric field strength for the corresponding target area (red-half-violin) and whole brain (blue-half-violins) separately. The split violin chart shows the results with DLPFC (A, C) and hippocampus (B, D) as the target region under tTIS (A, B) and tACS (C, D). 
 To explore the overall effect of stimulation mode and target selection  (tTIS vs tACS, DLPFC vs Hippocampus ) on subjects, we compared the mean field intensity of the whole-brain region after stimulation and conducted a two-way analysis of variance combined with individual differences between groups (E). *** represents $p < 0.001$. 
 	}
	\label{fig:violin}
\end{figure}

In this study, DLPFC and the hippocampus are selected as representative shallow and deep targets, respectively. We employ the MOVEA algorithm to obtain Pareto fronts that quantify the trade-off between intensity and focality for these two target regions. Fig.~\ref{fig:pareot_results} illustrates the individual differences in head shape and tissue morphology. Variability in tissue segmentation leads to inter-subject variability in the leadfield matrix, which in turn results in different Pareto fronts for each subject. For example, the maximum achievable intensity induced by tTIS targeting the DLPFC varies from 0.45 $V/m$ to 0.65 $V/m$.
Additionally, we measure the distribution of the electric field at each voxel when the target ROI achieves maximum intensity. In Fig.\ref{fig:violin}, when targeting the shallow region DLPFC (Fig.\ref{fig:violin}A), the electric field intensity within the target region (blue) exhibits a greater difference compared to the electric field intensity across the whole brain (red). Furthermore, we conduct a two-way analysis of variance (ANOVA) for eight subjects under the same simulation method and target, revealing statistically significant effects of simulation mode (p<0.001), individual differences (p<0.001), and the interaction between simulation mode and individual differences (p<0.001). Similar results are observed for the interaction between the simulation target and individual differences (p<0.001).

\section{Discussion}
In this paper, we propose MOVEA, an evolutionary algorithm-based model for solving multi-objective optimization in transcranial electrical stimulation. MOVEA provides a computational framework for optimizing the tES stimulation strategy without predefined stimulation parameters, obtaining optimal solutions as Pareto fronts with multiple constraints. MOVEA can further assist the comparison and selection of individual stimulation strategies, showing great potential in the application of personalized tES.

\textbf{Comparison with existing methods}: Our algorithm is capable of handling optimization for maximizing the magnitude of the electric field without specifying a predefined direction. While optimization for tACS without a predefined direction has been thoroughly explored \cite{saturnino2021optimizing,sadleir2012target}, the complexity of tTIS has led most current optimization methods to primarily focus on specific directions \cite{huang2020optimization}. In contrast to optimization problems with a prescribed direction as shown in Eq.\ref{eq:ti2}, optimization for maximizing the magnitude of the electric field requires consideration of interactions among electric fields along three orthogonal axes. Therefore, existing techniques, such as the current-flow model, which enables optimal solutions to be directly obtained from the leadfield matrix without algorithmic optimization \cite{huang2020optimization}, cannot be directly applied to maximum field optimization. Furthermore, in maximum field optimization, the vector $\vec{n}$ varies depending on the choice of activated channel, necessitating a more complex formula, as illustrated in Eq.\ref{eq:ti}. Designing a mathematical solver for this equation is challenging due to its complexity.

Here, our algorithm MOVEA employs an evolutionary algorithm, offering flexibility to adeptly address the inherent complexity of the problem. This flexibility is realized through the design of a decoding mechanism for the solution and the development of a suitable evolution strategy, both of which have been verified as effective. Stoupis et al. successfully employed an evolutionary algorithm to optimize focal stimulation in two-pair tTIS \cite{stoupis2022non}, utilizing the intensity-to-focality ratio  as a key evaluation metric. This ratio approach is commonly employed in existing heuristic methods involving multiple objectives, whether for focal stimulation \cite{stoupis2022non, bahn2021computational} or multi-target stimulation \cite{ruffini2014optimization, bahn2021computational}.
These methods, which do not search for specific solution values, require manual parameter adjustments and multiple runs to guide the algorithm towards the desired search direction. This includes adjusting the weight of the avoidance area \cite{saturnino2019simnibs}, the threshold for eliminating individuals \cite{stoupis2022non}, and the stringency of constraints \cite{huang2020optimization}. In addition to the challenges associated with manual parameter tuning, selecting optimal parameters necessitates comprehensive investigation, including defining parameter ranges and step sizes. Inappropriate adjustments may result in the inability to achieve a complete trade-off among competing objectives, a challenge that is further exacerbated when dealing with more than two objectives.

MOVEA employs Pareto optimization to treat each goal as an independent target, using Pareto non-dominance to connect them. After a single run, a comprehensive Pareto front is generated, containing a set of optimal solutions from which specific-value solutions can be chosen.

\textbf{Comparison of achieved intensity among different stimulation methods}:
Through the use of MOVEA, we discovered variations in maximum intensity achieved by different stimulation modalities. Specifically, two-pair tTIS yields the lowest maximum intensity, whereas HD-tTIS and HD-tACS achieve equivalent maximum intensities. This equivalence is observed only when HD-tTIS applies an identical montage to both frequencies and can be explained using Eq.~\ref{eq:ti2}.

When the desired direction aligns with the electric field itself, the electric field in the unconstrained direction can be represented using the directional electric field formula. Suppose the maximum field for HD-tACS is denoted as $\vec{E}{1}(\vec{r})\cdot (\vec{n})$. In the case of HD-tTIS, each individual electrode can transmit current at two distinct frequencies and intensities, resulting in two sub-fields, $\vec{E}{2}(\vec{r})\cdot (\vec{n})$ and $\vec{E}{3}(\vec{r})\cdot (\vec{n})$. The envelope is then calculated using Eq.~\ref{eq:ti2}. When the frequencies differ and the intensities are each equal to half of those for HD-tACS, the sub-fields are equivalent and correspond to $\vec{E}{1}(\vec{r})\cdot (\vec{n})/2$. In this scenario, the first term of Eq.~\ref{eq:ti2}, $| (\vec{E}{2}(\vec{r}) + \vec{E}{3}(\vec{r})) \cdot (\vec{n}) |$, reaches its maximum, while the second term, $| (\vec{E}{2}(\vec{r}) - \vec{E}{3}(\vec{r})) \cdot (\vec{n}) |$, becomes zero. Consequently, the envelope field attains its maximum value, $\vec{E}_{1}(\vec{r})\cdot (\vec{n})$, which is equal to the maximum field for HD-tACS. For the two-pair tTIS setting, hardware constraints make it infeasible for both pairs of electrodes to belong to the same channels. As a result, the second term of Eq.~\ref{eq:ti2} cannot be equal to zero, and the maximum intensity achievable by two-pair tTIS is lower than that achievable by HD electrode-based methods.

Based on the Pareto fronts obtained by MOVEA, we compared the performance of tES in terms of stimulation intensity (Fig.~\ref{fig:paretofrontier}\&\ref{fig:allrois}\&\ref{fig:pareot_results}). Consistent results regarding maximum intensity achieved by tES were observed across different head models, aligning with previous analyses. For example, HD-tTIS and HD-tACS achieve equivalent maximum intensities in the DLPFC region ($0.58 \text{V/m}$ for HD-tTIS/HD-tACS, and $0.56 \text{V/m}$ for two-pair tTIS), with the current injected for HD-tTIS at each frequency being precisely half that of HD-tACS. Our findings are congruent with previous studies~\cite{rampersad2019prospects,rampersad2019prospects,huang2019can,huang2021comparison}. For instance, Rampersad et al. simulated the intensity of two-pair tTIS and HD-tACS through an exhaustive search and found that the free-field intensity at the motor area under two-pair tTIS is $0.56 \text{V/m}$ and HD-tACS is $0.60 \text{V/m}$ \cite{rampersad2019prospects}.

\textbf{Focality and steerability of tTIS}: 
Intensity and focality represent conflicting objectives that cannot be simultaneously optimized to their fullest extent (Fig.\ref{fig:paretofrontier}\&\ref{fig:visualization}). In our evaluation of the focality performance of various stimulation modalities, we measured the intensity distribution across the entire brain while maintaining equivalent intensity levels at the target region. We found that HD-tTIS offers the best focality, followed by two-pair tTIS, with HD-tACS exhibiting the least focality. Fig.\ref{fig:paretofrontier} suggests that tTIS provides better focality than tACS at the same target intensity, as evidenced by the lower activation outside the target region induced by tTIS under the same conditions as tACS (Table~\ref{tab:activate_volume}). The superior focality of tTIS may be explained by its stimulation mechanism, in which only the low-frequency electrical signals obtained by coupling two high-frequency electrical inputs can effectively affect neural activity. This observation is further illustrated in Fig.~\ref{fig:visualization}, where two-pair tTIS generates relatively weak stimulations outside the target region (e.g., DLPFC and hippocampus). In contrast, tACS requires a greater number of electrodes to achieve maximum focality within the target region. HD-tTIS achieves optimal focality, but necessitates numerous activated channels with low currents ranging from 0 to 0.1 $mA$. when targeting a superficial region (e.g., V1) via tACS, the activation ratio $Vol$ is greater than that of a deeper region (e.g., Pallidum). The large activation ratio could be attributed to V1's proximity to the CSF. As suggested by Huang et al. \cite{huang2019can}, CSF is highly conductive, and targets closer to the CSF experience a more diffused electric field. Conversely, tTIS can effectively manipulate the activation ratio $Vol$ in deeper regions. As shown in Table~\ref{tab:activate_volume}, tTIS achieves a lower $Vol$ compared to tACS when targeting V1, resulting in a highly reduced ratio in $\Delta$, suggesting that tTIS can improve focality.

The results of MOVEA with an avoidance zone suggest that tTIS offers greater operability. Fig.~\ref{fig:study3} shows the stimulation intensity of tDCS, HD-tACS, two-pair tTIS, and HD-tTIS, with HD-tTIS outperforming the other methods. Traditional tACS, which employs large electrode patches, fails to generate satisfactory electric fields in the brain (Fig.~\ref{fig:study3}A).This focality problem can be improved with smaller electrodes, as suggested by Laakso et al. \cite{8105344}. Therefore, the performances of HD-tACS optimized by SimNIBS and MOVEA are better than traditional tACS (Fig.~\ref{fig:study3}B\&C). Moreover, because of the characteristics of tTIS, the activation of shallow areas can be mitigated but not entirely avoided when the electric field converges at the target. This property further improves the convergence and steerability, which is in line with previous study \cite{huang2019can,cao2019stimulus}.

\textbf{Inter-subject variability}: 
We evaluate the performance of MOVEA using head models from eight subjects and investigate the impact of inter-subject variability on the optimization of tES strategies. The variability in individual anatomical structures, such as skull thickness, CSF layout, and cortical folding, can lead to differences in the optimal stimulation strategies required to achieve maximum intensity at the target region~\cite{laakso2015inter}. This variability is demonstrated by the differences observed in the Pareto fronts shown in Fig.\ref{fig:pareot_results}, as well as the electric field strength of voxels at the target and whole brain under maximum intensity, as depicted in Fig.\ref{fig:violin}. Our findings are consistent with previous literature, which suggests that customized tES strategies, including tTIS and tACS, are necessary to enhance neuromodulation efficacy~\cite{lee2020individually,huang2021comparison}. Furthermore, the results of two-way analysis of variance (ANOVA) reveal significant differences in target selection, individual variability, and their interactions (Ps < 0.001), as well as significant differences between simulation modes and individual variability. These statistical results highlight the effectiveness of our target area and stimulus selection, and they reveal the interaction between these factors and individual variability.

Overall, despite the inter-subject variability, the effects of both tTIS and tACS are reliable, and the observed trends align well with existing studies. The ability to account for inter-subject variability further emphasizes the importance of individualized optimization in tES to address the inherent anatomical differences between individuals and to maximize the efficacy of neuromodulation~\cite{zhu2021optimization}.

\textbf{Limitations and future directions}:
While MOVEA offers significant advantages in providing a set of optimal solutions for flexible and personalized tES, there are several limitations that merit consideration:

Although MOVEA has achieved substantial improvements in computational efficiency compared to other genetic-based solvers, such as reducing the time required to obtain solutions from 2 hours for a single solution to 1 hour for an entire Pareto front~\cite{stoupis2022non}, it remains a time-consuming process. The most time-intensive step is calculating the electric field for millions of voxels distributed across the brain, which could limit its applicability in closed-loop neuromodulation scenarios. Future work will explore strategies to further improve efficiency by designing new metrics or optimization algorithms. For instance, model resolution could be reduced within regions of interest to accelerate the algorithm, and the electric field distribution outside the target area could be excluded from consideration. Then, our results reveal that there is a spatial discrepancy between the predefined target region and the location of the highest intensity point achieved through optimization. To address this challenge and improve the precision of stimulation, additional constraints or objectives could be introduced into the optimization process. For example, the half-maximum radius, a measure of the spread of the electric field around the maximum intensity point, could be optimized to ensure greater focus on the target region~\cite{dmochowski2011optimized}. Lastly, constructing personalized head models relies on the availability of individual MRI images, which may not always be feasible. Alternative methods that optimize tES without individual MRI have been proposed, such as using EEG signals to guide tES design~\cite{cancelli2016simple}. By integrating these approaches into MOVEA, the cost and complexity of personalized tES optimization could be reduced, making tES-based clinical treatments more accessible and feasible for real-world applications.

\section{Conclusion}
In conclusion, we have presented MOVEA, a versatile framework for optimizing tES policies. MOVEA is designed to accommodate multiple stimulation modalities, including tACS, HD-tACS, two-pair tTIS, and HD-tTIS. Additionally, MOVEA is compatible with user-defined constraints, such as the number of electrodes and the desired intensity within an avoidance region. MOVEA obtains a set of solutions at each level of objective (\eg different intensities), avoiding the situation that specific solutions cannot be obtained due to individual differences.
Through the application of MOVEA, we evaluated and compared the performance of various stimulation modalities. Our analysis revealed that both HD-tACS and HD-tTIS are capable of achieving higher stimulation intensities. Furthermore, among the modalities examined, HD-tTIS stands out for its superior focality, making it particularly effective in precisely targeting specific brain regions.
MOVEA provides personalized tES strategies with customized constraints, offering a potential tool for clinical treatments and uncovering the causal relationship among brain area, cognitive function, and behavior.

\section{Declaration of competing interest}
The authors declare that they have no known competing financial interests or personal relationships that could have appeared to influence the work reported in this paper.

\section{Declaration of Generative AI and AI-assisted technologies in the writing process}
Statement: During the preparation of this work the authors used ChatGPT in order to check grammar and refine the language. After using this tool/service, the authors reviewed and edited the content as needed and take full responsibility for the content of the publication.

\section*{Acknowledgments}

This work was funded in part by the National Key R\&D Program of China (2021YFF1200804), National Natural Science Foundation of China (62001205), Shenzhen Science and Technology Innovation Committee (20200925155957004, KCXFZ2020122117340001, JCYJ20220818100213029), Shenzhen-Hong Kong-Macao Science and Technology Innovation Project (SGDX2020110309280100), Guangdong Provincial Key Laboratory of Advanced Biomaterials (2022B1212010003).

\bibliographystyle{plain}  
\bibliography{reference.bib}

\end{document}


\section{Pseudocode}

\subsection{objective function}
\subsubsection{tACS}
\begin{equation} 
E_\Omega = \frac{1}{G_\Omega} \int_{\Omega} \bm{{\Vert E \Vert}}{d}G = \frac{1}{G_\Omega} \int_{\Omega} \sqrt{w_xE_x^{2}+w_yE_y^{2}+w_zE_z^{2}}{d}G,
\label{eq:tACS-intensity}
\end{equation}
where $G_\Omega$ denotes the total volume in the brain area $\Omega$. $x$, $y$ and $z$ represent three orthogonal directions, and $w_x$, $w_y$, and $w_z$ denote the weights assigned to the respective orthogonal directions.

\subsubsection{tTIS}
The envelope field in the brain without a predefined direction can be calculated as Eq.~\eqref{eq:ti}:
\begin{equation}
\left|\vec{E}_{AM}(\vec{r})\right|=\left\{\begin{array}{l}
 2\left|\vec{E}_{2}(\vec{r})\right|,   \qquad\qquad\qquad\qquad \qquad\qquad \: |\vec{E}_{1}(\vec{r})| < |\vec{E}_{2}(\vec{r})| cos(\alpha)\\
2\left|\vec{E}_{2}(\vec{r}) \times\left(\vec{E}_{1}(\vec{r})-\vec{E}_{2}(\vec{r})\right)\right| /\left|\vec{E}_{1}(\vec{r})-\vec{E}_{2}(\vec{r})\right|, \: otherwise
\end{array}\right.
\label{eq:ti}
\end{equation}

where $\vec{E}_{1}(\vec{r})$ and $\vec{E}_{2}(\vec{r})$ are the first and second distinct fields at the location $\vec{r}$ and $\alpha$ is the angle between $\vec{E}_{1} $ and $ \vec{E}_{2} $. The Eq.~\ref{eq:ti} holds only when $\alpha$ < 90° and $|\vec{E}_{1}(\vec{r})| > |\vec{E}_{2}(\vec{r})|$, or the sign of $ \vec{E}_{2} $ and the numbering of the channels should be inverted and swapped respectively.

\subsection{Pseudocode}
\subsubsection{Stage\_One}
\begin{algorithm}
\caption{Stage\_One}\label{Stage_{One}}
\SetKwInOut{Input}{input}\SetKwInOut{Output}{output}
\Input{ size of population : $population\_size$, coordinate of target : $Coordinate$, maximum iteration : $iteration1$ }
\Output{the solution with maximum fitness: $optimal$}
Initialize individuals as$population\_size$ randomly \;
\While{($iteration1 > 0$)}{
Select individuals for reproduction randomly\;
Perform crossover and mutation to produce offspring\;
Evaluate the fitness of the offspring by $Coordinate$ and Objective function Eq.(\ref{eq:tACS-intensity} \& \ref{eq:ti})\;
Replace the least fit individuals in the population with the offspring\;
$iteration1 \leftarrow iteration1 - 1$\;
}
Evaluate the fitness of each individual in the population\;
\end{algorithm}

\newpage
\subsubsection{Stage\_Two}
\begin{algorithm}[h]
\caption{Stage\_Two}\label{Stage_{Two}}
\SetKwInOut{Input}{input}\SetKwInOut{Output}{output}
\Input{size of population : $number\_of\_particles$,size of archive : $size\_of\_archive$;coordinate of target : $coordinate$, maximum iteration : $iteration2$, the result from : $Stage\_One$ $optimal$}
\Output{the Pareto front : $archive$}
    Set position for one particle by $optimal$ and initialize the others randomly in the search space\;
    initialize velocity for all particles randomly\;
    \While{($iteration2 > 0$)}{
    Calculate the objective functions for each particle by $coor$, Objective function Eq. (\ref{eq:tACS-intensity} \& \ref{eq:ti}), constraints Eq.~\eqref{con4} \; 
    Calculate the Pareto front by Algorithm~\ref{evaluate_solution} and add it to the $archive$\;
    \If{size of the $archive$ $ > size\_of\_archive$ }{
        Delete superfluous solutions by Algorithm~\ref{crowding} \;
    }
    Update globe optimal solution from $archive$ \;
    Update the $velocity$ and $position$ of each particle by Eq. (\ref{eq:velocity-update} \& \ref{eq:position-update}) \;
    Update personal best position\;
    $iteration2 \leftarrow iteration2 - 1$\;
}
\end{algorithm}

\subsubsection{Main Functions}
\begin{algorithm}[h]
\caption{Initiation}\label{initiation}
\SetKwInOut{Input}{input}\SetKwInOut{Output}{output}
\Input{the size of population : $size\_of\_population$, the size of archive : $size\_of\_archive$, the solution from $Stage\_One$ : $optimal$, sparse constant : $sparse$ }
\Output{population : $[position,velocity]$}

$position$ of one particle $ \leftarrow optimal$;

\For{$position \leftarrow$ other particles}{ 
    \For{$dimension \leftarrow position$}{
        $dimension \leftarrow$ uniformly random from -$sparse$ to $sparse$\;
        \If{$dimension > 1 $ or $ dimension < -1 $}{
            $dimension \leftarrow 0$
        }
    }
 }   
initialize $velocity$ for all particles from $0$ to $1$ randomly\;
\end{algorithm}

\begin{algorithm}
\caption{Crowding\_Distance\_Sorting}\label{crowding}
\SetKwInOut{Input}{input}\SetKwInOut{Output}{output}
\Input{ population : $population$, the size of archive : $size\_of\_archive$ }
\Output{new population : $new\_population$}
rank $population$ to Pareto fronts$F$ based on Algorithm~\ref{evaluate_solution}\;
$i \leftarrow 0$ \;
$new\_population \leftarrow [ ]$ \;
\While{$|new\_population| + F_i < size\_of\_archive$}{
$i \leftarrow i + 1$\;
add $F_i$ to $new\_population$ \;}
\If{$|new\_population| < size\_of\_archive$}{
  sort the solutions in $F_i$ according to one goal\;
  the crowding distance of the first and last solutions in $F_i \leftarrow$ infinity\; 
  \While{ $|new\_population| < size\_of\_archive$}{
    add the solution with the maximum crowding distance to $new\_population$ and remove it from $F_i$\;}}
\end{algorithm}

\begin{algorithm}
\caption{Evaluate\_solutions}\label{evaluate_solution}
\SetKwInOut{Input}{input}\SetKwInOut{Output}{output}
\Input{constraint violation value for solution $i$ : $CV(i)$, $m$-th fitness value of  solution $i$ : $f_m(i)$, intensity of target by solution $i$ : $f_1(i)$, a large constant : $penalty$}
\Output{Pareto fronts $Pareto\_Fronts$}

\For{$i\leftarrow$ solution from $population$}{
\uIf{$CV(i)$ is not zero }{ 
  \For{$m\leftarrow$ 1 to $M$}{ $f_m(i) \leftarrow CV(i) \cdot penalty$\; }  }
\ElseIf{$f_1(x) < 0.1$ }{
      \For{$m\leftarrow$ 1 to $M$}{ $f_m(i) \leftarrow f_m(i) \cdot penalty$\;} }
}
Get $Pareto\_Fronts$ based on $f(i)$ fast nondominated sort\; 
\end{algorithm}